\documentclass[preprint]{aastex}

\newcommand{\HI}{H$\,${\sc i}}
\newcommand{\HII}{H$\,${\sc ii}}
\shortauthors{Iyer et al.}
\begin{document}

\title{Neutral Hydrogen in Arp 158}

\author{Mansie G. Iyer and Caroline E. Simpson}
\affil{Department of Physics, Florida International University, Miami, Fl 33199}
\email{miyer01@fiu.edu, simpsonc@fiu.edu}

\author{Stephen T. Gottesman }
\affil{Department of Astronomy, University of Florida, Gainesville, FL 32611}
\email{gott@astro.ufl.edu}

\and

\author{Benjamin K. Malphrus }
\affil{Department of Physical Sciences, Morehead State University, Morehead, KY 40351}
\email{b.malphrus@morehead-st.edu}
\begin{abstract}

We present 21 cm observations of Arp 158. We have performed a study of the neutral hydrogen (\HI) to help us understand the overall formation and evolution of this system. 
This is a disturbed system with distinct optical knots connected by a linear structure embedded in luminous material. There is also a diffuse spray to the southeast. The \HI\ seems to be made up of three distinct, kinematically separate systems. Arp 158 bears a certain optical resemblance to NGC 520 (Arp 157), which has been identified as a mid-stage merger.
 From our 21 cm observations of Arp 158, we also see a comparable \HI\ content with NGC 520. These similarities suggest that Arp 158 is also an intermediate stage merger.

\end{abstract}


\section{INTRODUCTION}

In the early seventies, Toomre (1970) and Toomre \& Toomre (1972), presented numerical models which demonstrated that 
strong tidal forces between interacting galaxies could result in features like 
plumes, shells, rings, tidal tails, and bridges. 
They also proposed 
that such strong collisions between galaxies would lead to orbital decay and eventual
merging. In 1977, Toomre identified a series of galaxies that he believed 
represented galaxies at different stages of merging (``The Toomre Sequence'') and proposed that 
the end product of such merging could be an elliptical
galaxy. This hypothesis came to be known as the ``merger hypothesis.''

Hibbard \& van Gorkom (1996) presented H-$\alpha$, R-band, and 21 cm observations of several 
galaxy systems from the Toomre sequence and showed that these galaxies most likely do form an evolutionary 
chain. Three stages in the merging sequence were identified: (i) the early stage, in which the 
disks are well separated and only marginally disrupted, (ii) the intermediate stage, which exhibits
distinct nuclei embedded in luminous material, and (iii) the late stage, in which tidal appendages are 
seen to emerge from a single nucleus.

The purpose of this study is to better understand and interpret the evolutionary sequence as described by Hibbard \& van Gorkom (1996). This project was initially part of a multiwavelength project, in collaboration with H. Deeg and C. Mu\~{n}oz-Tu\~{n}\'{o}n of the Instituto de Astrofisica de Canarias in Spain (Deeg et al. 1998). We have picked seven galaxies: Arp 213, Arp 78, Arp 135, Arp 31, Arp 279, Arp 263 and Arp 158, which we believe are in various stages of the evolutionary sequence. In this first paper we report on neutral hydrogen (\HI) observations of Arp 158.

Arp 158 (NGC 523) is located at $\alpha$ = 01$^{h}$ 25$^{m}$ 20.8$^{s}$ and  $\delta$ = + 34\arcdeg 01$\arcmin$ 29$\arcsec$ (J2000).
It is described in Zwicky's Catalogue of Selected Compact Galaxies and Post-Eruptive Galaxies (1971) as  
``post-eruptive, blue compact with three compact knots connected by a bright bar, fan-shaped jets and matrix.'' 
The characteristics of the galaxy are given in Table 1.
Arp's image (Figure 1) shows that there are three visible knots, with a central linear structure. (Without appropriate kinematic information, it is impossible to determine whether this is a bar.) There is also a bright tail extending to the west and an extensive, faint diffuse tail to the south-east. Only the western and eastern knots are easily visible in this image; the third knot is located approximately in the center of the linear structure.

\placetable{Table1}

\placefigure{fig1}

The westernmost ``nucleus'' was examined by Chincarini \& Heckathorn (1973, hereafter referred to as C\&H) and was claimed to be a foreground star. From a close visual exam of Arp's image, this knot appears to be an unresolved point source, and resembles other stars in the image. Dahari (1985), however, argued that his measured redshift for this knot was similar to that of the south-east one and thus could not be a foreground star. Without confirming spectra, we cannot determine the nature of this knot, or establish its connection with Arp 158. For the purposes of discussion, we will refer to it as the western knot/star.

\section{OBSERVATIONS AND DATA REDUCTION}

High sensitivity observations of the 21 cm emission of Arp 158 were made during May 14 and May 17, 1999  by M. Iyer,  
 C. Simpson and B. Malphrus with the Very Large Array\footnote{The Very Large Array is operated by the National Radio Astronomy Observatory (NRAO), which is a facility of the National Science Foundation operated under the cooperative agreement by Associated Universities, Inc.} (VLA) in its spectral line mode. The galaxy was observed with the D-array configuration for a total of five hours, including move time and calibration time, 
 using a 128 channel spectrometer with a total bandwidth of 6.25 Mhz. On-line hanning smoothing was not used, so the velocity resolution is 1.2 times the channel separation (Wilcots et. al, 1997). For our observations, with a channel separation of 48.8 kHz (10.3 km s$^{-1}$), this results in a velocity resolution of 12.3 km s$^{-1}$. The observations were made at a
central velocity of 4758 km s$^{-1}$ and the data were collected for a single polarization. The observational parameters are given in Table 2.
 
\placetable{Table2}
 
Standard calibration and editing procedures were performed using the AIPS data reduction package available from the NRAO. The 
continuum sources 0119+321 (J2000) and 0137+331 (J2000) were used to calibrate the phase and amplitude response of the receiving system. The data
from each day were edited and calibrated independently before being combined in the \textit{uv}-plane to produce the combined data set. 

The data reduction for the May 17 data set was complicated owing to  solar interference. The D-array configuration is very susceptible to solar contamination, particularly at short baselines. To rectify this problem, all the data points from baselines less than 1k$\lambda$ (210 m) for the May 17 data were not used in the imaging process. 
 This was approximately 13\% of the total data for both the days combined and reduces by $\approx$ 7\% the sensitivity of our maps, especially to large scale structures. Narrow band interference was also present in channels 99 and 100 but as these channels were not included in the calibration and did not include any line emission, they did not interfere with our reduction or analysis.

After calibration, the AIPS task UVLIN (Cornwell et. al, 1992) was used to subtract the continuum emission. The task makes linear fits to the visibility data in a range of channels without \HI\ line emission and then subtracts the requisite values from all the channels. In our dataset, channels 6 to 33 (5376 km s$^{-1}$ and 5088 km s$^{-1}$) and channels 102 to 119 (4354 km s$^{-1}$ and 4174 km s$^{-1}$) were line-free and were used to perform the continuum subtraction.
A fast-Fourier transform was then applied to this \textit{uv} data to produce a cube ($\alpha$, $\delta$, v) of images, one at each observed frequency (velocity channel). During imaging, each channel  was CLEANed (H\"{o}gbom 1974; Clark 1980) to reduce the effect of the sidelobes produced by non-gaussian features of the synthesized beam (the ``dirty'' beam). We tested various weighting schemes (robust parameter in the IMAGR task in AIPS) during the imaging process to increase the resolution, but because of the lack of baselines less than 1k$\lambda$, the increase in resolution was not worth the corresponding decrease in sensitivity. The higher resolution images did not reveal any new morphological or kinematic features, and with the lowered sensitivity, low column density features were lost completely. Hence, in order to maximize the sensitivity to low surface brightness features, we used the natural weighting scheme, resulting in a beamsize (FWHM) of 47.3 $\arcsec$  $\times$ 42.2 $\arcsec$.
 The data cube was cleaned down to 0.98 mJy per beam, corresponding to 1$\sigma$ as determined by the statistical analysis
of a signal-free channel map. 
Line emission signal was present in channels from 41 to 89, approximately 4500 km s$^{-1}$ to 5000 km s$^{-1}$, which were then integrated to produce moment maps representing the \HI\ integrated column density (0th moment), the temperature-weighted mean velocity (1st moment), and the velocity dispersion (2nd moment).

\section{Results}

\subsection{\HI\ Spatial Distribution, Column Densities and Masses}

Throughout this paper we use a value of H$_0$ equal to 75 km s$^{-1}$ Mpc$^{-1}$ and a heliocentric velocity of 4758 km s$^{-1}$ ( NASA IPAC Extragalactic Database, NED\footnote{http://nedwww.ipac.caltech.edu}) for Arp 158. These values imply a distance of 63 Mpc. The results from the derived quantities are shown in Table 3, and are discussed below.

\placetable{Table3}

Figure 2 is a grayscale image of the \HI\ integrated flux density. 
 We can see that there are three distinct \HI\ concentrations which we denote as A, B, and C. Figure 3 represents the optical image from the second generation  blue Digitized Sky Survey (DSS2)\footnote{http://skyview.gsfc.nasa.gov} overlaid with \HI\ contours. The central optical knot is more easily visible in this image. 

\placefigure{fig2}

\placefigure{fig3}

For the entire distribution, we have detected a total \HI\ mass of 6.5 $\times$ 10$^{9}$ M$_\odot$. System A, which has the highest beam-smoothed \HI\ column density (N$_{HI}$ = 13.0 $\times$ 10$^{20}$ atoms cm$^{-2}$) coincides with the optical knot system and seems to be centered to the west of the western knot/star. System B (N$_{HI}$ = 7.4 $\times$ 10$^{20}$ atoms cm$^{-2}$) appears to coincide with the diffuse tail to the south-east. There is also a very interesting gas knot to the north-west of the galaxy. This is system C (N$_{HI}$ = 1.2 $\times$10$^{20}$ atoms cm$^{-2}$). Note that the field of view for Arp's image (Figure 1) does not include this area. Although this system is faint in \HI\, it  does appear in four contiguous channels and we believe it to be a valid detection.
Using separate integration maps, we have calculated \HI\ masses associated with each of our defined systems (see section on kinematics), as well as their flux-weighted systemic velocities (Table 3), and an estimate of their velocity widths (FWHM). System A has the highest \HI\ mass (2.9$\times$ 10$^{9}$ M$_\odot$) and the \HI\ masses associated with systems B and C are approximately 1.9 $\times$ 10$^{9}$ M$_\odot$ and 0.1 $\times$ 10$^{9}$ M$_\odot$, respectively.

Using the optical angular size (2.5$\arcmin$ $\times$ 0.7$\arcmin$) catalogued in NED, the optical dimensions for Arp 158 are 46 $\times$ 13 kpc. The total \HI\ distribution is spread well beyond the optical image. We measure an \HI\ extent of 70 $\times$ 33 kpc (not deconvolved), measured out to 2 $\times$ 10$^{19}$ atoms cm$^{-2}$.

Looking at Figure 3, there is faint optical emission located north and very slightly east of the \HI. The northernmost part of the \HI\ contour might overlay the southernmost part of the optical emission, but the \HI\ is not obviously coincident with this weak optical emission. A deeper optical image is needed to establish its nature and a spectrum would be required to establish any physical connection to Arp 158.

\subsection{Kinematics}

Figure 4 shows the intensity-weighted velocity map. The isovelocity contours are very different for systems A and B, and there is an area of steep gradient between the two which overlies the bright optical emission in the system. There also appear to be three distinct kinematic areas coinciding with systems A, B, and C. The apparent kinematical axes of systems A and B, as indicated by the isovelocity contours, are oriented in different directions. The position angle of the kinematic axis for system A, measured from north through east, is approximately 25\arcdeg\ on the western side, rotating to a position angle of approximately 90\arcdeg\ on the eastern side. This almost 70\arcdeg\ kink in the isovels in system A shows that this is a fairly abrupt change. For system B, the position angle is approximately 150\arcdeg.

\placefigure{fig4}

The channel maps (Figure 5) for Arp 158 reveal that there is emission in every channel between systems A and B, showing that there is gas extending between them which is superposed along the line-of-sight in the moment maps.
The extended emission could be either a bridge or a tail.
It could be that systems A and B were two different galaxies that are now interacting. In this case, we could consider the extended emission to be a bridge. We do not believe this to be the case. Optically, systems A and B do not have separate nuclei. The distinct optical knots only seem to coincide with system A, whereas system B does not have much of a stellar component associated with it.

\placefigure{fig5}

It seems more likely that B is a tidal tail that formed due to the interaction that deformed the optical components of the system.  
The center point in each of the channel maps is moving smoothly to the west going to higher velocities. The contours for system B show a smooth run in velocity, with a few widely separated isovels indicating a shallow observed gradient in velocity. It also exhibits a sharp cut-off in velocity space at the westernmost end. Thus system B, which optically resembles a tail, also stretches out in the \HI\ and has a narrow spread in velocity. All these properties are consistent with this feature being a tidal tail (Hibbard \& van Gorkom, 1996), perhaps seen edge-on, leading to the superposition of gas we see in the moment maps. However, the shallow velocity gradient we observe could also be due to gas moving mostly transverse to our line of sight.

Because of the superposition of different kinematic components of this complex object, we have used the channel maps in conjunction with position-velocity diagrams (Figure 6) to define (somewhat subjective) velocity ranges for systems A, B, and C, as well as for the gas ``tail'' connecting A and B. To produce these position-velocity diagrams, we used the AIPS task XSUM to (a) sum along the declination at the central right ascension and (b) sum along the right ascension at the central declination. Using our velocity ranges, we then integrated the appropriate channel maps for each of these components separately to produce moment maps for each object and the gas tail.

\placefigure{fig6}

The velocity field for system A is shown in Figure 7a, which was created by integrating only the channels from 4503.0 km s$^{-1}$ to 4694.2 km s$^{-1}$ in the cleaned cube. System C, which overlaps with A in velocity space, appears in the channel maps from 4566.7 - 4598.6 km s$^{-1}$. Figure 7b, showing only system B, was created by integrating the cube from 4907.0 km s$^{-1}$ to 5002.8 km s$^{-1}$. Figure 7c, showing the connecting gas tail, is from the integration of 4704.8 km s$^{-1}$ to 4885.7 km s$^{-1}$ . For all the three parts of Figure 7, the isovels are 10 km s$^{-1}$ apart. The kinematic axes of systems A and B are more clearly visible in these figures, and are oriented in different directions. The isovels in the connecting tail appear to broaden moving from system A southeastward toward system B.

\placefigure{fig7}

Using our \HI\ flux-weighted systemic velocites, we find that systems A and B are separated by almost 343 km s$^{-1}$, whereas systems A and C are separated by about 33 km s$^{-1}$. The velocity widths (Figure 8) for systems A, B and C are approximately 256 km s$^{-1}$, 76 km s$^{-1}$ and 30 km s$^{-1}$. The velocity widths as well as the \HI\ masses were measured using only the channels defined for each object. Except for system C, which is clearly separate, please note that these quantities are dependent upon our velocity ranges for each object. 

\placefigure{fig8}

\subsection{Continuum Emission}

There is an unresolved continuum source, NVSS J012521+340128 (Condon et al. 1998), at the position of Arp 158 (Figure 9), with a catalogued flux density of 15.4 mJy. It is located in the area between systems A and B, roughly coincident with the start of the connecting gas tail. 
This could either be a background source, unconnected to the Arp 158 system, or a source within the system. As it is not uncommon for continuum emission to be associated in some way with interacting systems, it is highly likely that the continuum source is connected with Arp 158. The idea that radio continuum emission was related to galaxies undergoing an interaction was initially conceived due to the connection of three radio sources with optically strange galaxies (Baade \& Minkowski, 1954).

\placefigure{fig9}

In either case, there is a possibility of absorption of the continuum emission by the \HI\ in Arp 158. As only the line-free channels were used to estimate the continuum intensity for subtraction from our data cube, we have confidence that our continuum subtraction was not affected by any absorption that may be present. 
There could however, be an issue with the measured \HI\ intensity at the position of the source. If the level is low between systems A and B, it could be that there  really is less \HI\, or that there in a reduction in the \HI\ intensity due to absorption. There is nothing obvious in the spectrum of the peak pixel of the continuum source to indicate the presence of absorption, and the integrated flux of the continuum source made from channels with and without line emission were comparable.

We also attempted to check for absorption in the \HI\ line at the location of the continuum source by comparing an ``off-source'' spectrum through the cube to an ``on-source'' spectrum. The on-source spectrum is run at the position of the continuum source, while the off-source spectrum is obtained from an area of emission away from the continuum source. The two spectra are then subtracted, and if absorption is present, an absorption line profile may then be revealed (England \& Gottesman 1990). However, the success of this method depends on the two spectra being from kinematically similar areas, since the off-source spectrum is supposed to represent what the \HI\ line ``should'' look like (for no absorption). Unfortunately, even the closest off-source spectra we looked at are located in areas with \HI\ kinematics that are very different than the kinematics in the on-source area. This prevents us from directly determining the presence or absence of absorption at the location of the continuum source. However, since the measured intensity of the continuum source was the same in maps made with and without \HI\ emission, any absorption would have to be less than a few times the noise in our channel maps. This would indicate that the \HI\ is optically thin.

\section{Discussion}

In the optical image for Arp 158, there are bright components which include the knots and the linear structure, as well as a diffuse component (the south-eastern tail). 
There is still debate on the true identity of the westernmost nucleus, i.e. whether or not it is a foreground star.
Through this discussion, however, we assume the westernmost nucleus to be a foreground star as suggested by C\&H.

On the basis of the optical and \HI\ data, we believe Arp 158 to be an intermediate stage merger, consistent with the model proposed by Hibbard \& van Gorkom (1996).
Using Arp's image, we estimate that the knots and the linear structure extend to approximately 27 kpc. The bright tail, defined as the central knot to the beginning of the diffuse tail, is about 12 kpc. The extensive, faint diffuse tail to the south-east spreads out approximately another 26 kpc. From their electrograph at the Cassegrain focus of the 82 inch Struve reflector at the McDonald Observatory, C\&H's measurements are fairly consistent with ours. They estimated the optical extent of the bright components to be approximately 21 kpc, with the diffuse components extending to about 41--52 kpc. There are a few small optical condensations in the faint south-east tail; they may be \HII\ regions (C\&H), which would not be unusual in a tidal tail (Hibbard \& van Gorkom 1996).

C\&H presumed that the eastern knot and the central knot were two separate galactic nuclei from two interacting galaxies. We have estimated their projected separation to be approximately 6 kpc. Bernl\"{o}hr (1993) calculated a difference of 100 km s$^{-1}$ between the optical eastern and central knots, which is in agreement with that of C\&H. The eastern and central knots lie within the east part of the \HI\ system A and the west part of the connecting ``tail.'' From their spectra, C\&H assumed a similar composition and mass-to-light ratio for the knots, and attributed the optical brightness of the eastern knot to a higher central star density. Using this information, they calculated the mass ratio of the two nuclei to be 5:1. This  mass ratio implies that the gravitational interaction between the two nuclei is weak. To get substantial disruption in a weak encounter, the objects need to pass very close to each other in a prograde encounter. Since Arp 158 is very disrupted, the encounter must have been close. According to A. Toomre as reported by C\&H, the best proof that such an encounter had occurred was the optical tail in the south-east direction. 

By asssuming the galaxy to be completely spherical and calculating the ratio of the major and minor axes of the central knot, C\&H estimated Arp 158's inclination at 25\arcdeg. If this is a reasonable estimate, then the system is seen somewhat edge-on. Although this inclination estimate cannot be considered very accurate since the system is disturbed, the flat distribution of the \HI\ in our images is further indication that Arp 158 is probably an edge-on system. The superposition of the tail is expected if the system is edge-on and/or the plane of interaction is along the line of sight. This superposition is clearly seen in both the channel maps and the position-velocity diagrams. The low spatial resolution of our data precludes any in-depth inclination analysis, so the question of the true nature of its \HI\ morphology remains a question.

Based on our \HI\ data for Arp 158, system B, associated with the faint diffuse south-east optical extension, and the \HI\ tail together have \HI\ features that resemble a tidal feature. It is not clear if system C is, or has been, involved in this interaction. We do not see a stellar component associated with system C nor is it very well resolved. System C could be a tidal dwarf formed as a result of the interaction, a nearby galaxy, or perhaps a counterarm to the ``tail.'' The lack of any apparent gas connecting it to the rest of the system would argue against it being a counterarm. On the other hand, its narrow spread in velocity and its location, almost directly opposite system B, are consistent with system C being a counterarm.

Toomre also noted the optical similarity between NGC 520 (Arp 157) and Arp 158 (C\&H).
NGC 520, classified as an intermediate stage merger by Hibbard \& van Gorkom (1996), consists of two parent nuclei separated by $\approx$ 6 kpc (from images kindly supplied by J. Hibbard). There is also a tail to the south which extends to $\approx$ 23 kpc. There is a plume which emanates from this tail, bends east, and then stretches to the dwarf galaxy UGC 957 at a distance of $\approx$ 52 kpc. Like NGC 520, Arp 158 shows distinct nuclei immersed in luminous material, which is consistent with an intermediate stage merger. In addition, the tails in the two systems extend to almost the same length and the separation between the parent nuclei for NGC 520 and the separation between the eastern and central nuclei for Arp 158 are comparable. Furthermore, the \HI\ for Arp 158 extends beyond the optical image, and this is also true for NGC 520. The total HI mass for NGC 520 was measured to be 7.1 $\times$10$^{9}$ M$_\odot$ (Hibbard \& van Gorkom 1996), roughly similar to our measurement of Arp 158, which is 6.5 $\times$10$^{9}$ M$_\odot$.

\section{SUMMARY}

We believe Arp 158 to be a mid-stage merger. Optically, it has a very disturbed morphology, with bright knots connected by a linear structure, and a faint diffuse tail. The \HI\ is spread out far beyond the optical image, which is not unusual for interacting systems. There are three distinct HI concentrations, which are kinematically quite different. System A coincides with the optical knot system and system B appears to coincide with the diffuse south-east tail. System B also stretches out in the \HI\, and with its narrow spread in velocity, has features coincident with those of a tidal tail. System C is a previously undetected entity. There is no obvious optical counterpart to this \HI\ system, other than weak optical emission located just north of it. It would require a deeper optical image and spectra to verify any association of the optical emission with system C.

There is a continuum source in the vicinity of the galaxy. The westernmost optical knot also needs to be studied further to determine whether it is associated with Arp 158 or if it is a foreground star. There are distinctive similarities between Arp 158 and NGC 520, identified as a mid-stage merger, which lends credence to our belief that Arp 158 is also an intermediate stage merger. 

This research has made use of the NASA/IPAC Extragalactic Database (NED) which is operated by the Jet Propulsion Laboratory, Calafornia Institute of Technology, under contract with the National Aeronautics and Space Administration.

This work has also made use of NASA's \textit{Skyview} facility\footnote{http://skyview.gsfc.nasa.gov} located at NASA Goddard Space Flight Center to obtain optical images from the second Digitized Sky Survey. The Second Palomar Observatory Sky Survey (POSS-II) was made by the California Institute of Technology with funds from the National Science Foundation, the National Aeronautics and Space Administration, the National Geographic Society, the Sloan Foundation, the Samuel Oschin Foundation, and the Eastman Kodak Corporation.

The compressed files of the ``Palomar Observatory - Space Telescope Science Institute Digital Sky Survey'' of the northern sky, based on scans of the Second Palomar Sky Survey are copyright (c) 1993-2000 by the California Institute of Technology and are distributed herein by agreement. All Rights Reserved.
Produced under Contract No. NAS5-26555 with the National Aeronautics and Space Administration.

The authors would like to thank the referee, John Hibbard, for his very useful comments and suggestions on this paper.

\pagebreak

\clearpage
\pagebreak



\begin{figure}
\plotone{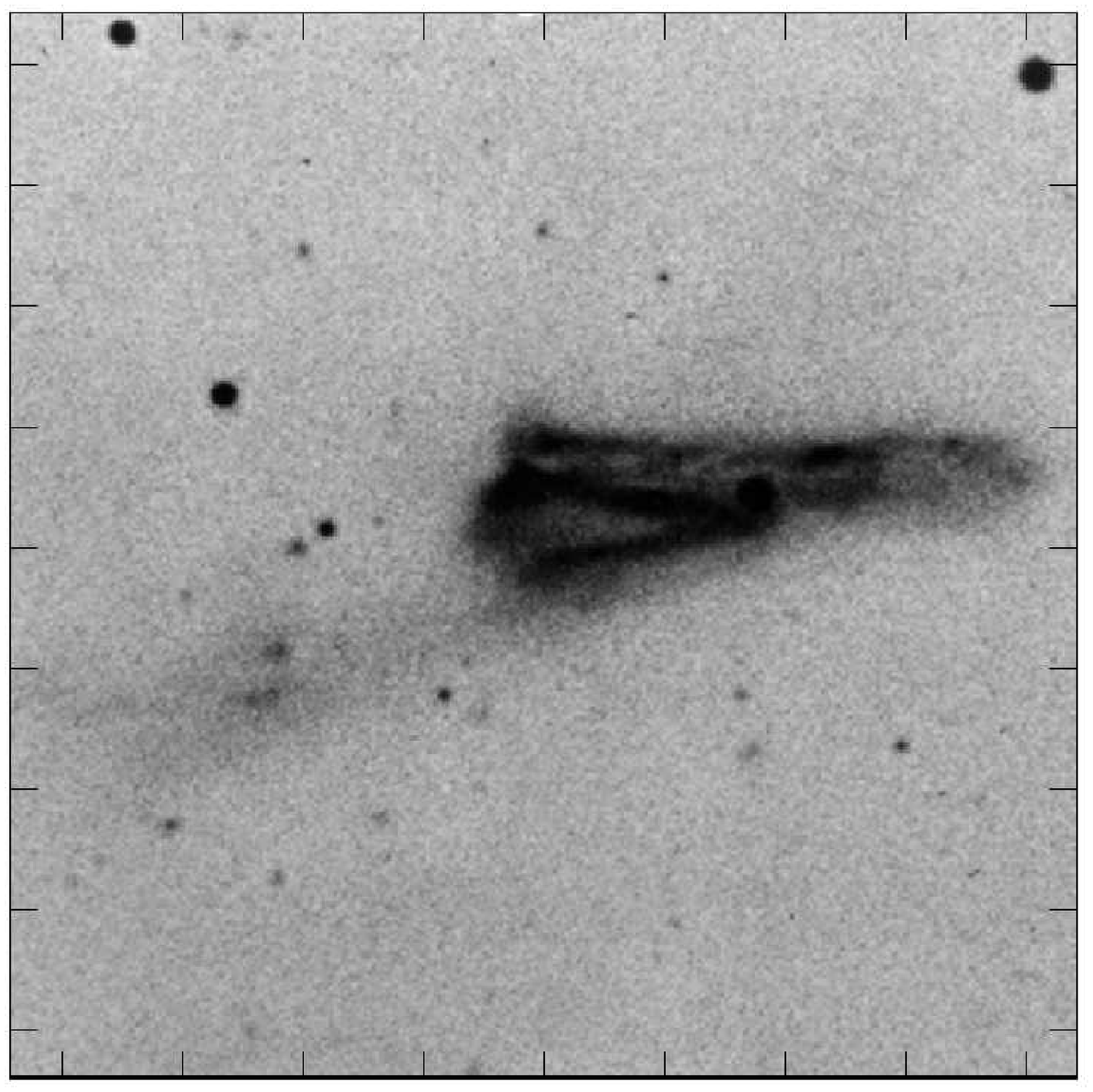}
\caption{Optical image of Arp 158 from Arp's Catalogue (1966), north is up; east is to the left. The photographic emulsion 103a-D was used to obtain the image.}
\label{Fig1}
\end{figure}

\begin{figure}
\plotone{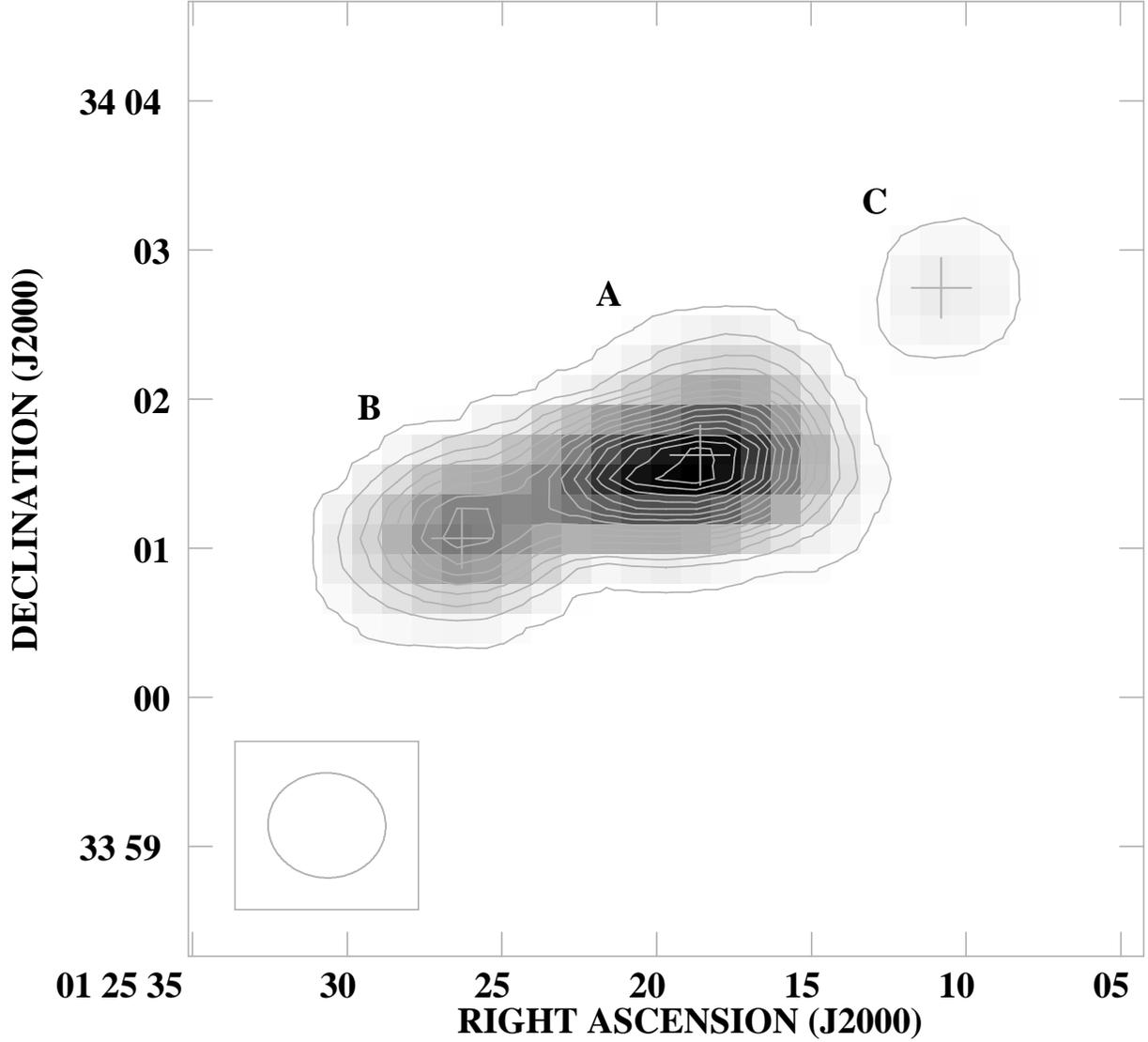}
\caption{Integrated \HI\ for Arp 158.
The contour levels are (2, 12, 22, 32, 42, 52, 62, 72, 82, 92, 102, 112, 122, 132, 142) $\times$ 10$^{19}$ atoms cm$^{-2}$, where 10$^{19}$ atoms cm$^{-2}$ = 18.09 Jy beam$^{-1}$ m s$^{-1}$. 
The letters A, B and C represent the three
different gas components of Arp 158. The enscribed ellipse in this figure and the following moment maps represents the beam size (FWHM = 47.3$\arcsec$ $\times$ 42.2$\arcsec$ ). \label{Fig2}}
\end{figure}

\begin{figure}
\plotone{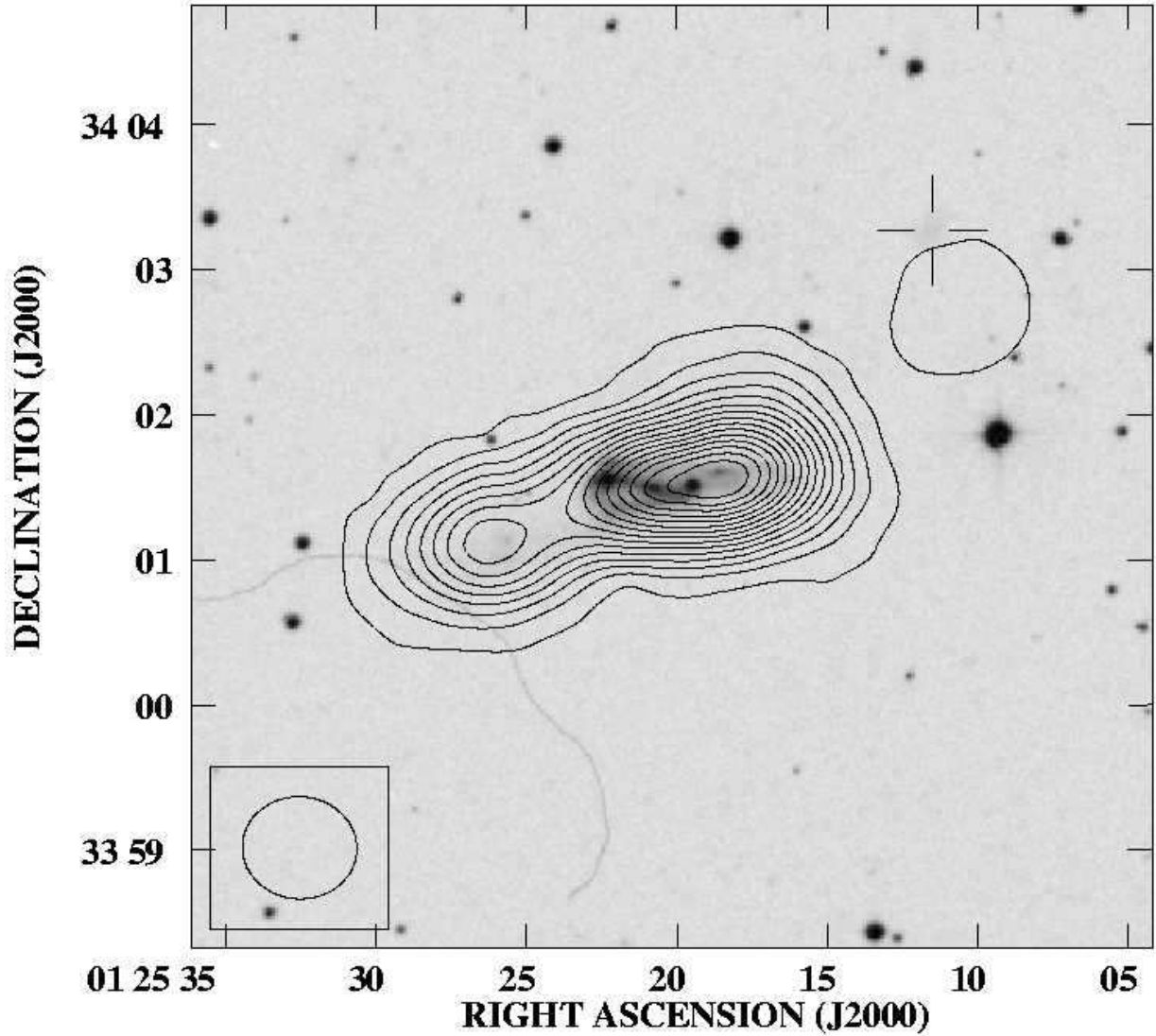}
\caption{The blue DSS2 optical image of Arp 158 with integrated \HI\ flux density contours. The contour levels are those used in Figure 2. The cross represents the position of the weak optical emission.\label{Fig3}}
\end{figure}

\begin{figure}
\plotone{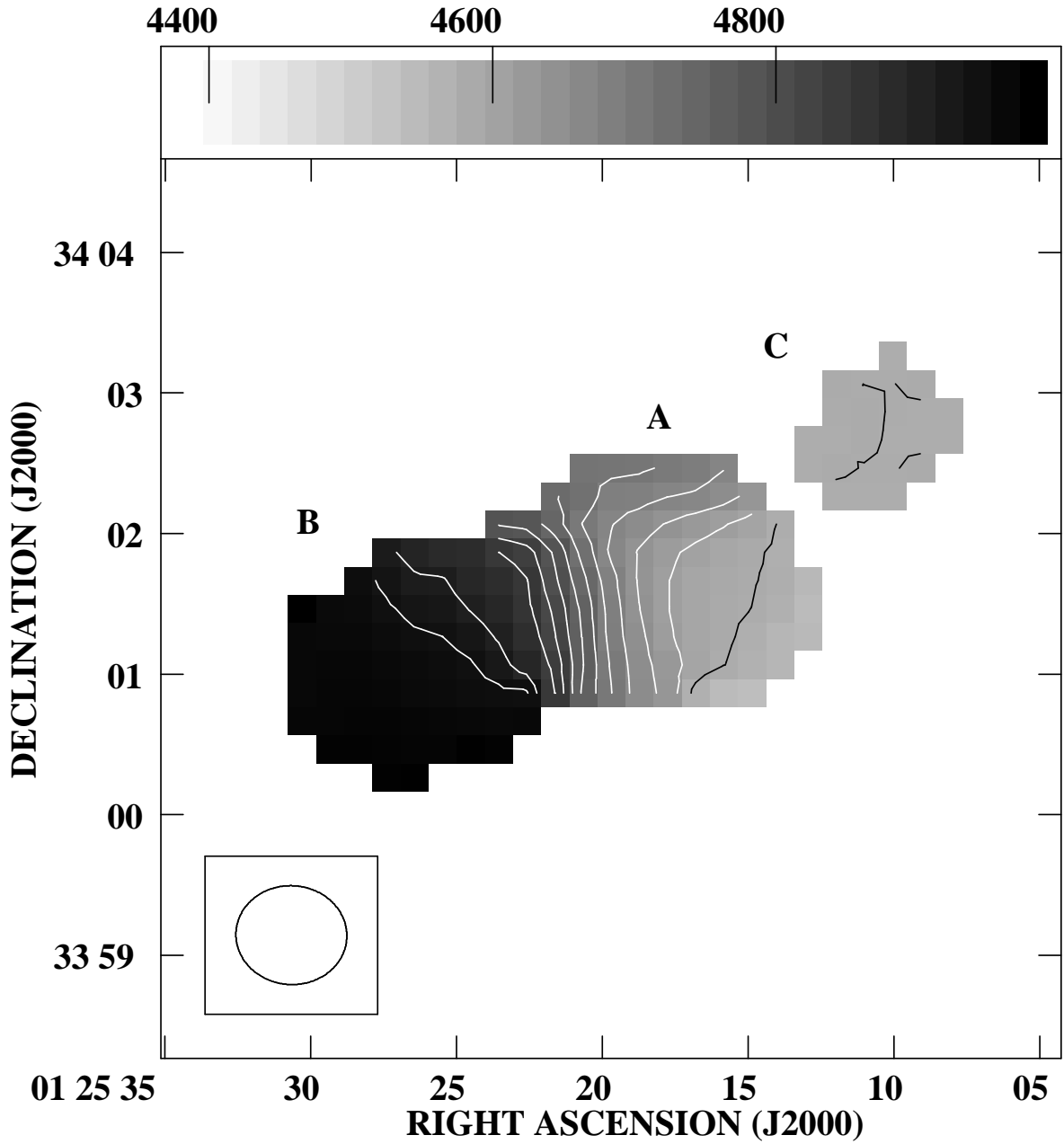}
\caption{Temperature weighted mean velocity map of Arp 158. Contours are plotted every 30 km s$^{-1}$ from 4580 to 4940 km s$^{-1}$.\label{Fig4}} 
\end{figure}

\begin{figure}
\figurenum{5a}
\epsscale{0.8}
\plotone{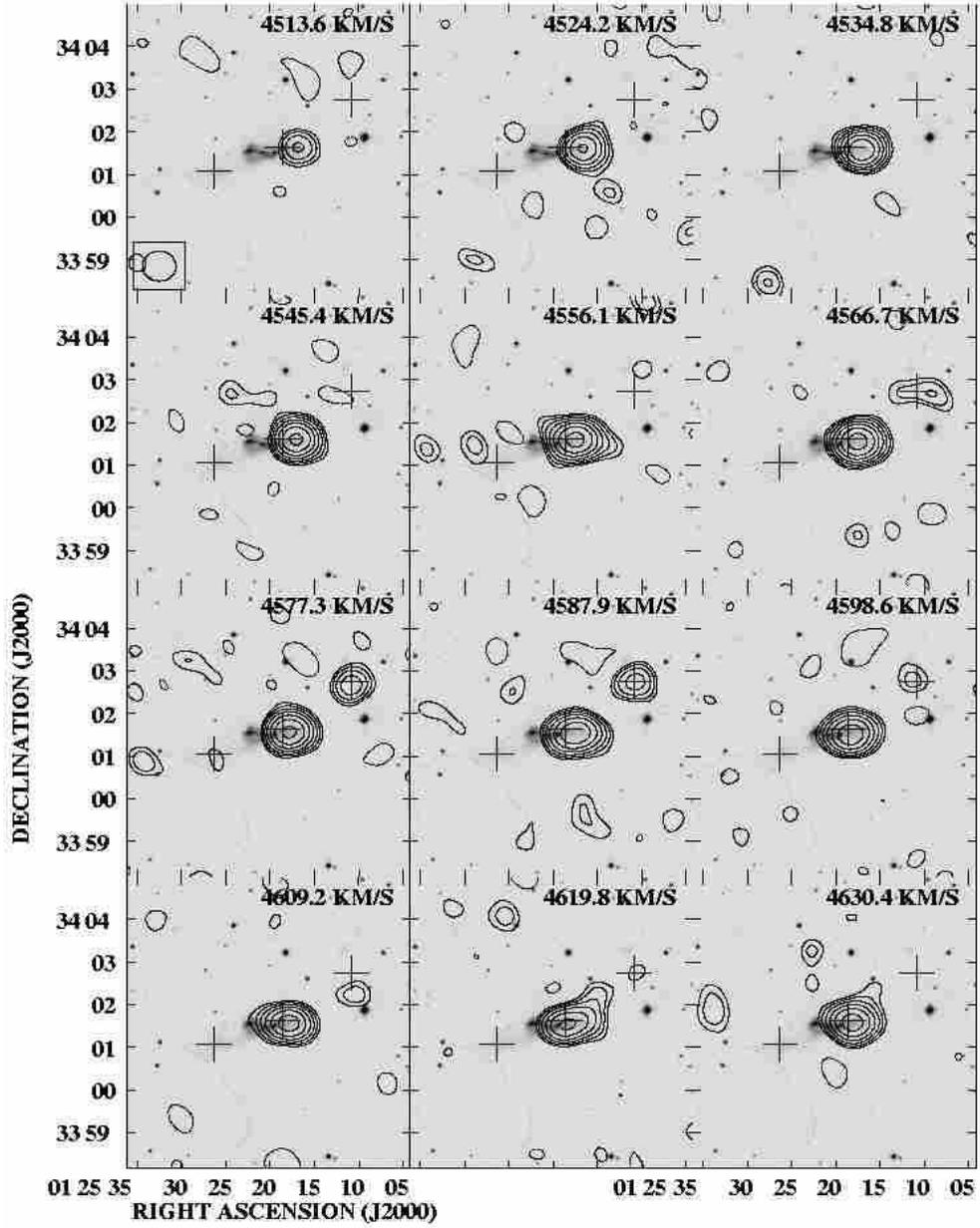}
\caption{\HI\ channel maps of Arp 158 are contoured on the optical blue DSS2 image. The heliocentric velocity for each channel is listed in the upper right corner. The crosses represent systems C, A, and B, from upper right to lower left respectively. Contours are at -2$\sigma$, 2$\sigma$, and then increase by $\sqrt{2}$. 1$\sigma$ = 0.98 mJy beam$^{-1}$. The highest contour is at 15.52 mJy beam$^{-1}$ and the dashed lines represent negative contours.
\label{Fig5a}} 
\end{figure}

\begin{figure}
\figurenum{5b} \plotone{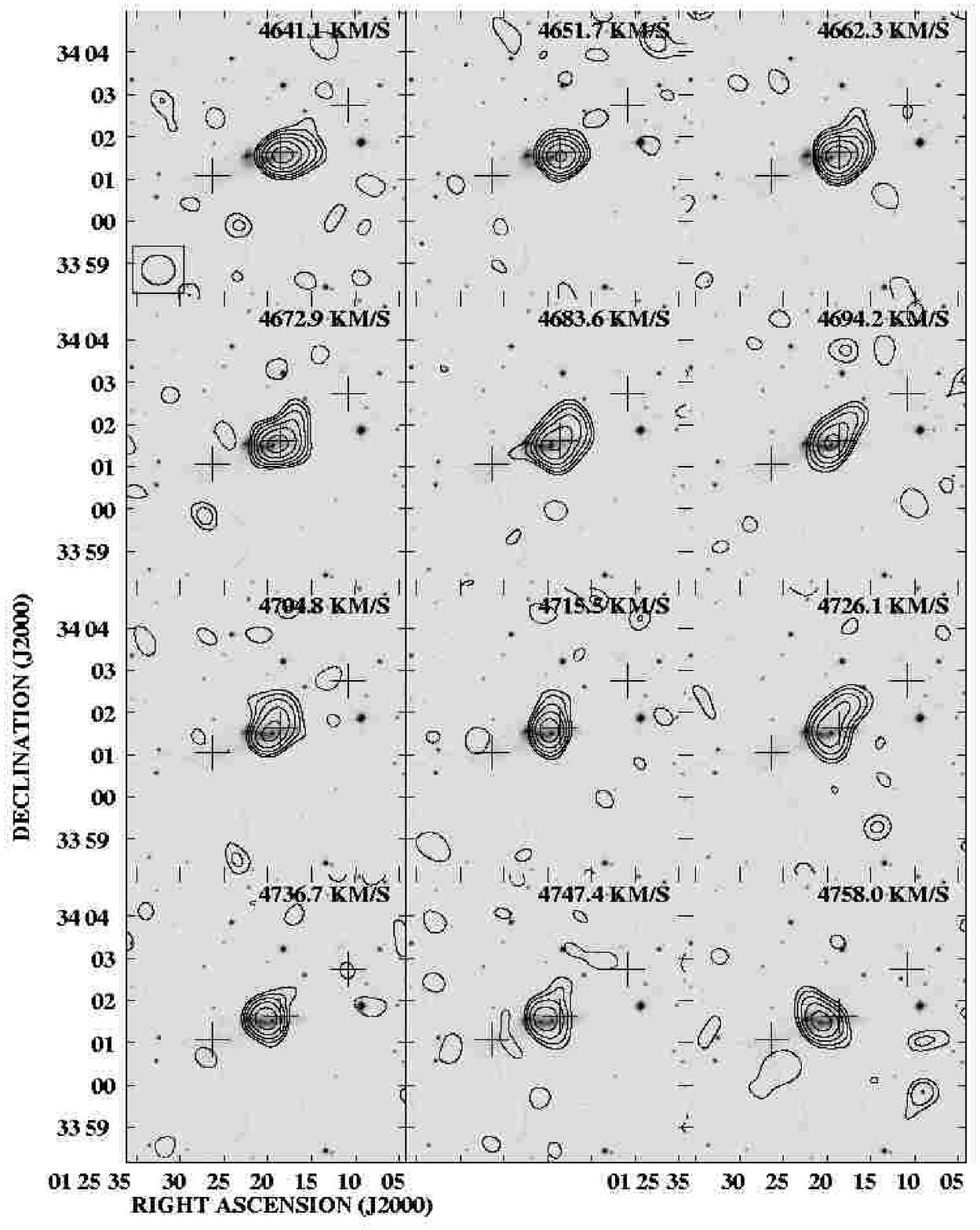}
\caption{Channel Maps (continued).\label{Fig5b}}
\end{figure}

\begin{figure}
\figurenum{5c}
\plotone{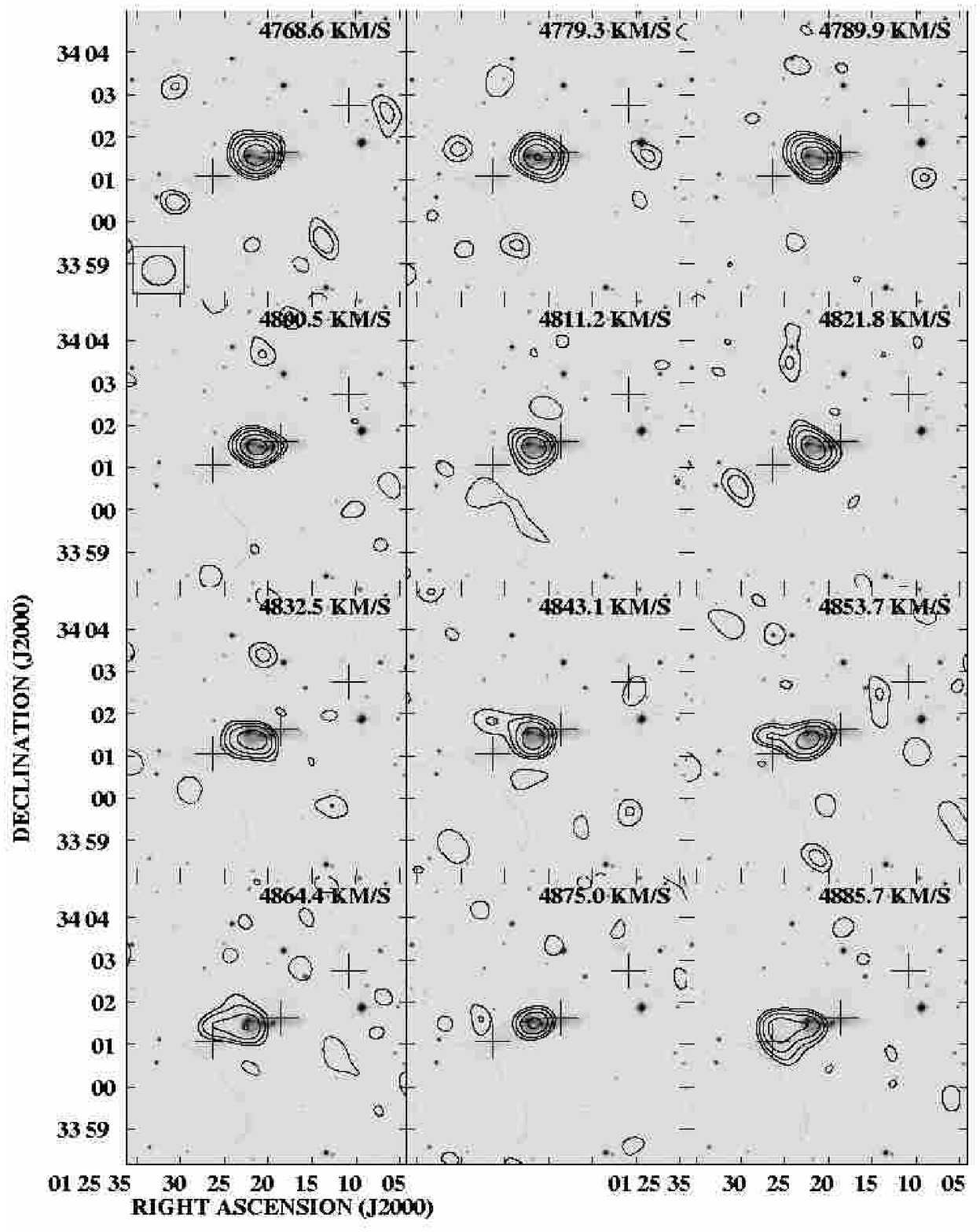}
\caption{Channel Maps (continued).\label{Fig5c}}
\end{figure}

\begin{figure}
\figurenum{5d}
\plotone{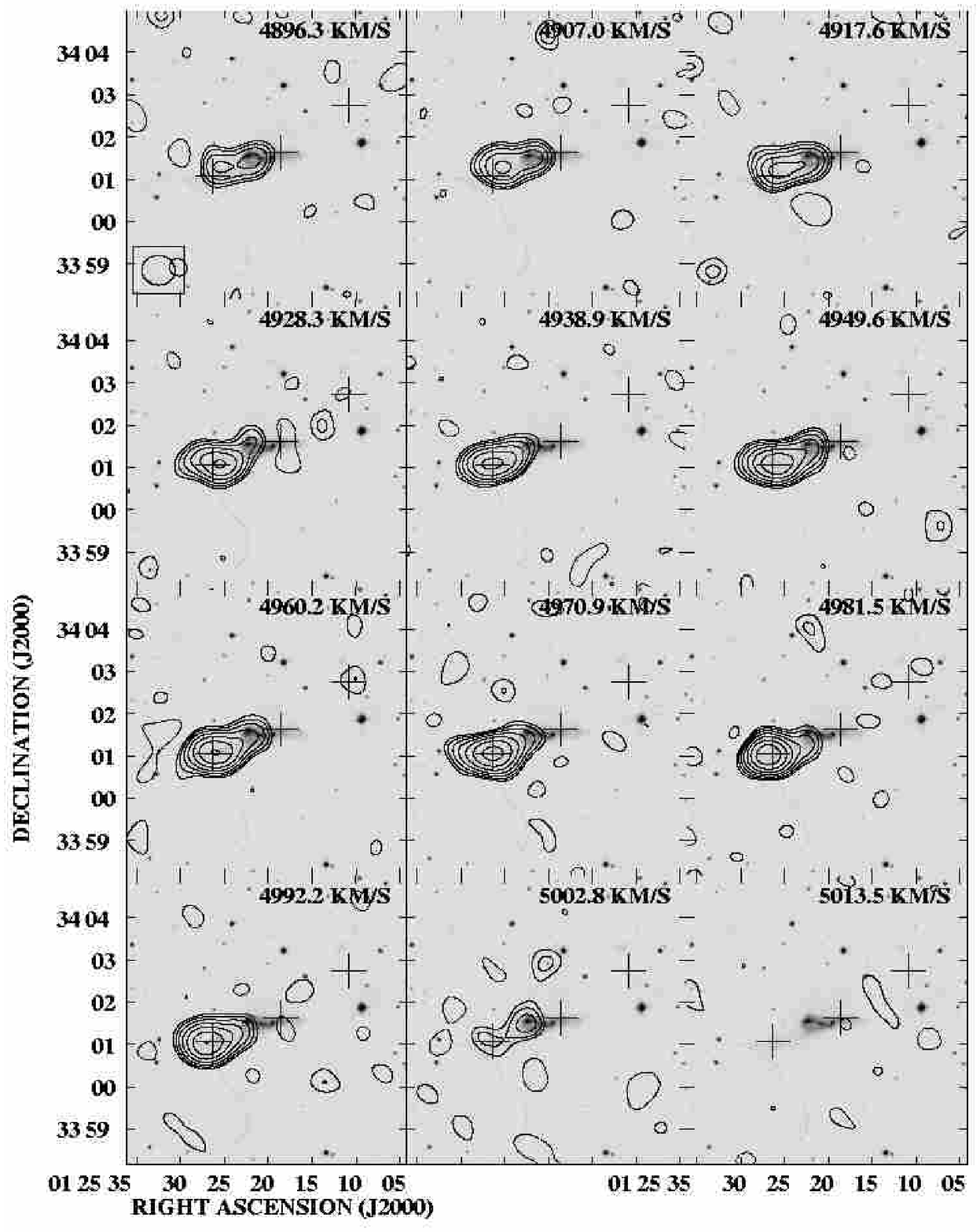}
\caption{Channel Maps (continued).\label{Fig5d}}
\end{figure}

\begin{figure}
\figurenum{6a}
\plotone{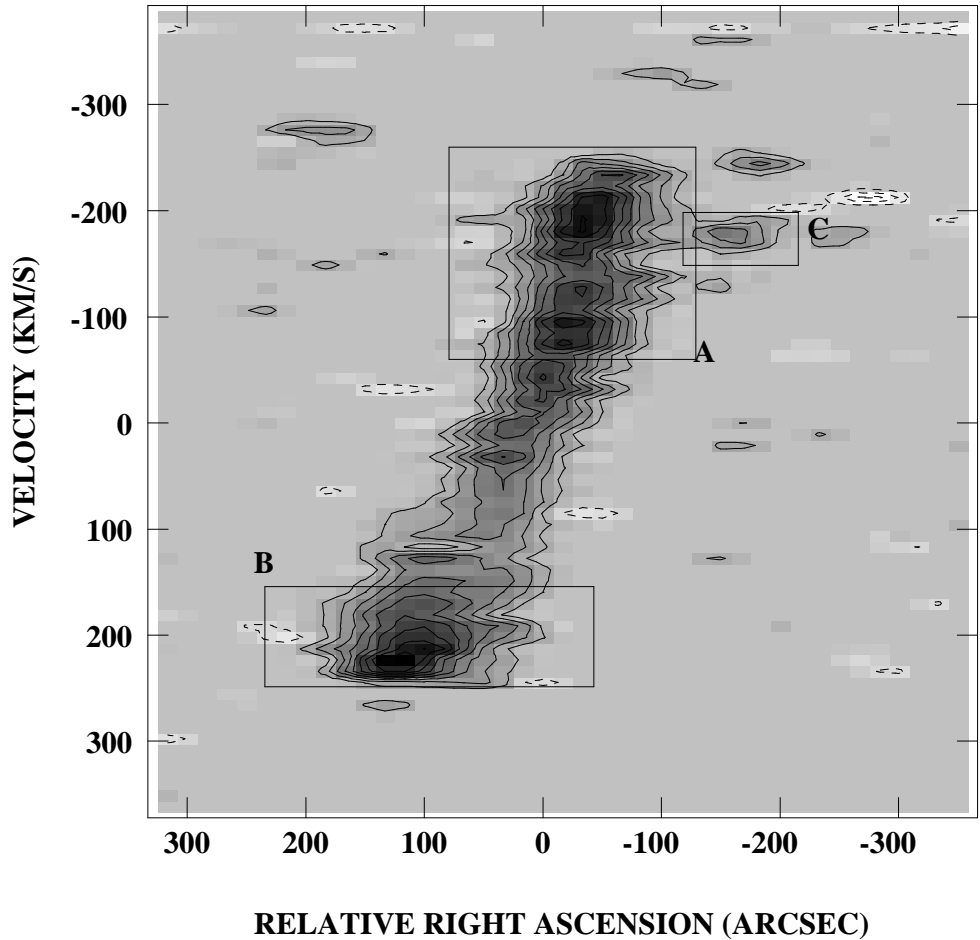}
\caption{Position-Velocity diagrams for Arp 158. Velocity relative to the systemic velocity (4758 km s$^{-1}$) vs. right ascension relative to the center of the field (01$^{h}$25$^{m}$19.7$^{s}$). Contours are plotted at (-100, -90, ...,+90, +100) \% of the peak flux, which is 0.078 Jy beam$^{-1}$.\label{Fig6a}}
\end{figure}

\begin{figure}
\figurenum{6b}
\plotone{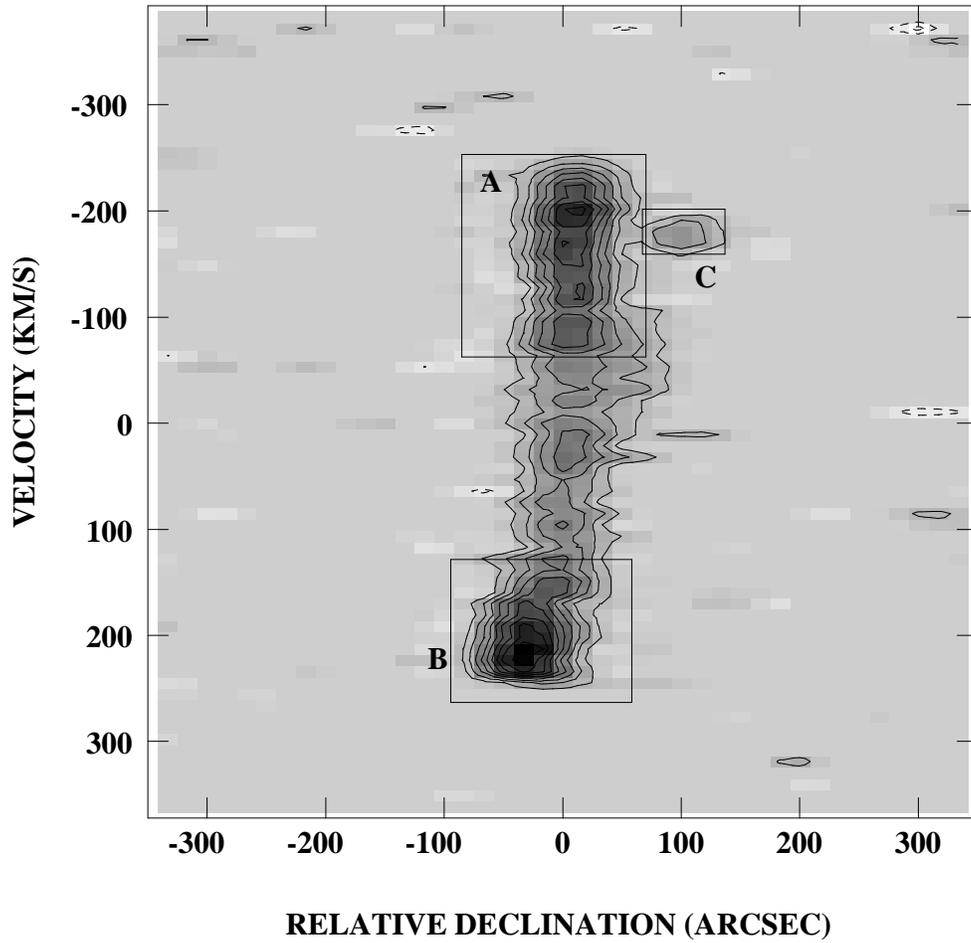}
\caption {Declination relative to the center of the field (34\arcdeg01$\arcmin$28$\arcsec$) vs. velocity relative to the systemic velocity. The contours are percentages of the peak flux, as for (a); however in this case, the peak flux is 0.105 Jy beam$^{-1}$. The boxes in each of these figures approximately represent the velocity ranges used for producing the moment maps in Figure 7.\label{Fig6b}}
\end{figure}

\begin{figure}
\figurenum{7a}
\plotone{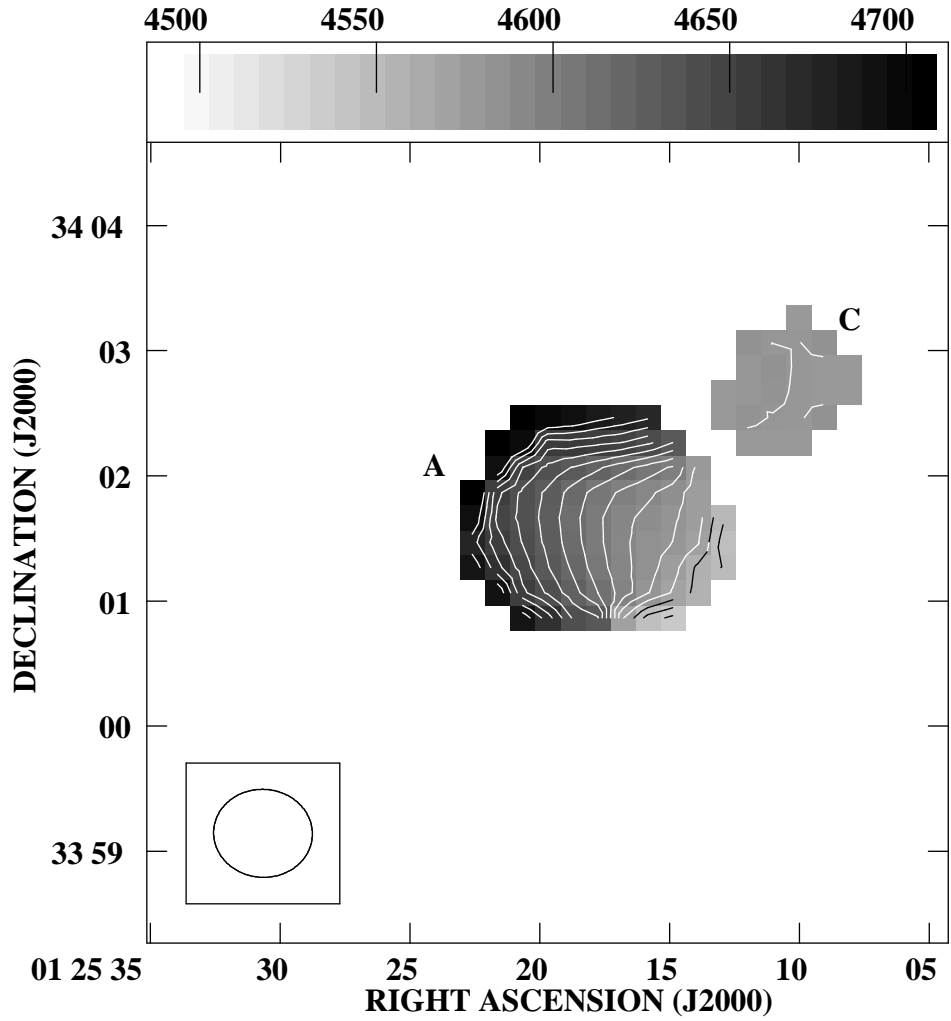}
\caption{Temperature-weighted mean velocity maps of Systems A and C. The contours are plotted every 10 km s$^{-1}$ from 4540 to 4680 km s$^{-1}$.\label{Fig7a}}
\end{figure}

\begin{figure}
\figurenum{7b}
\plotone{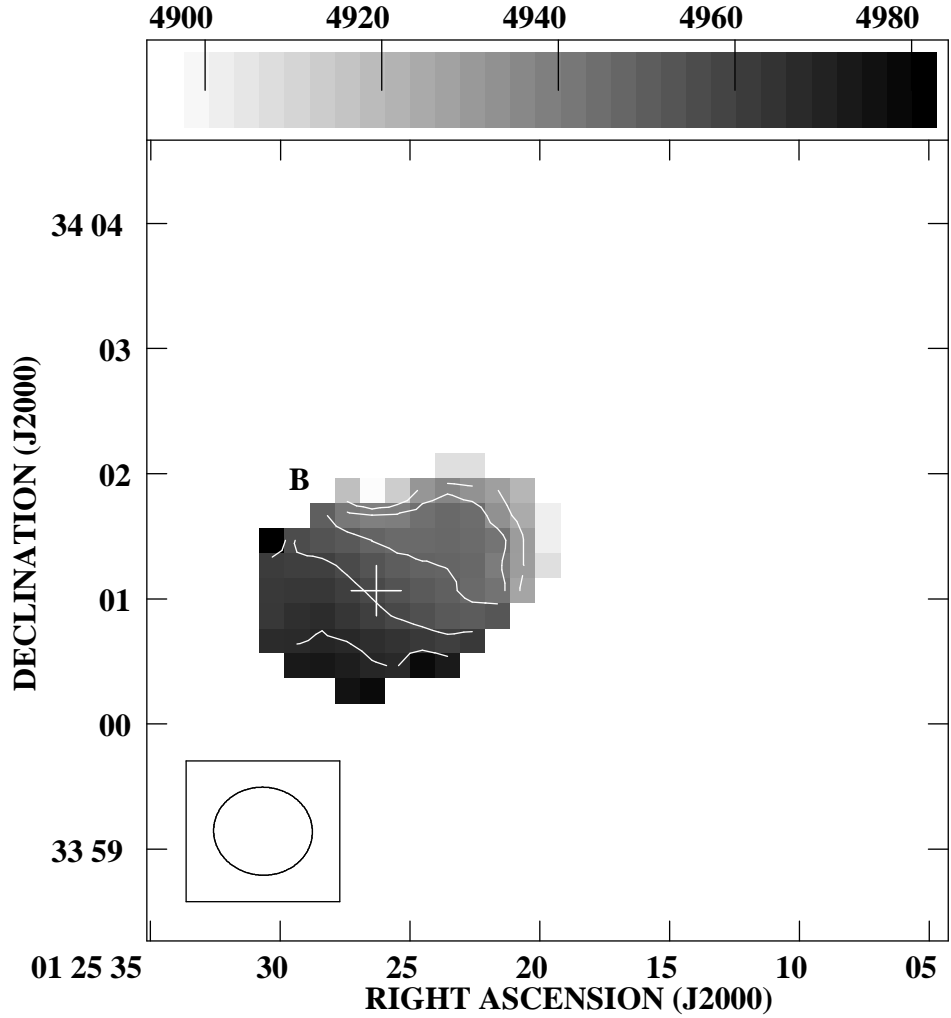}
\caption{Temperature-weighted mean velocity map of system B. The contours are plotted every 10 km s$^{-1}$ from 4928 to 4968 km s$^{-1}$. \label{Fig7b}}
\end{figure}

\begin{figure}
\figurenum{7c}
\plotone{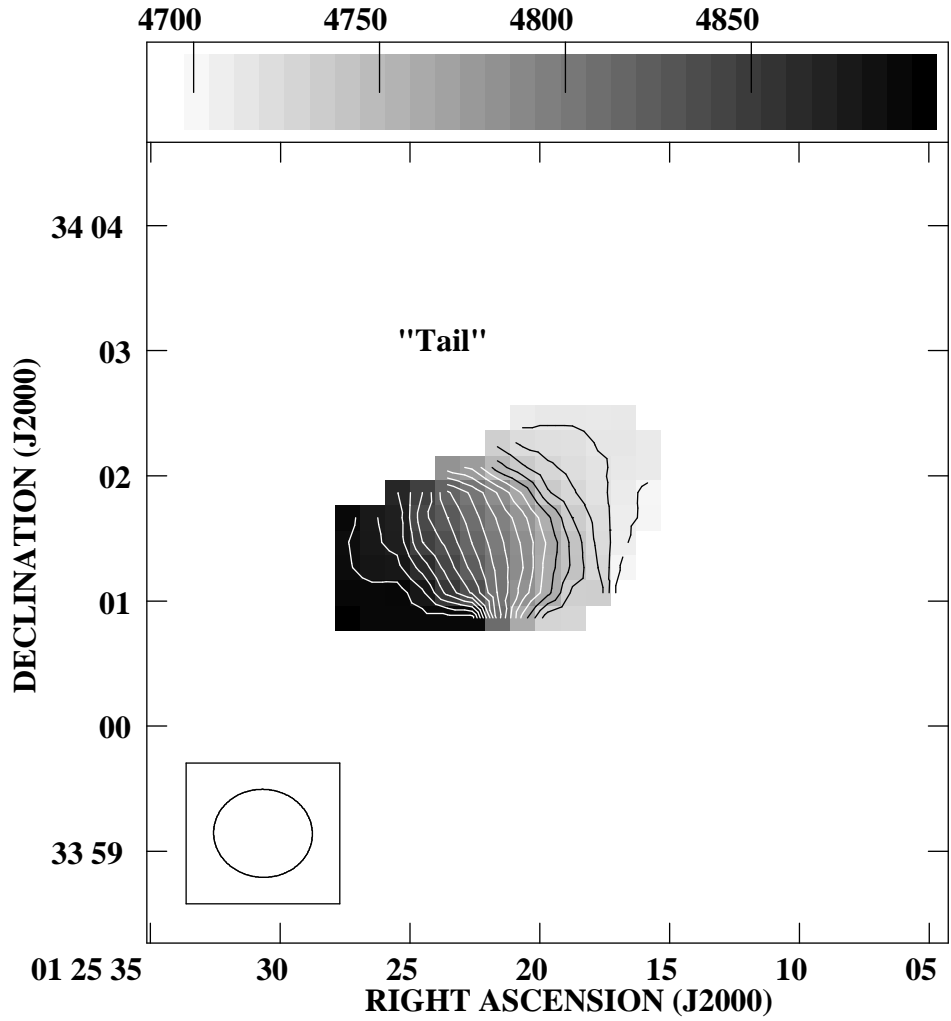}
\caption {Temperature-weighted mean velocity map of the ``tail'' connecting systems A and B. The contours are plotted every 10 km s$^{-1}$ from 4705 to 4885 km s$^{-1}$.\label{fig7c}}
\end{figure}
 

\begin{figure}
\figurenum{8}
\plotone{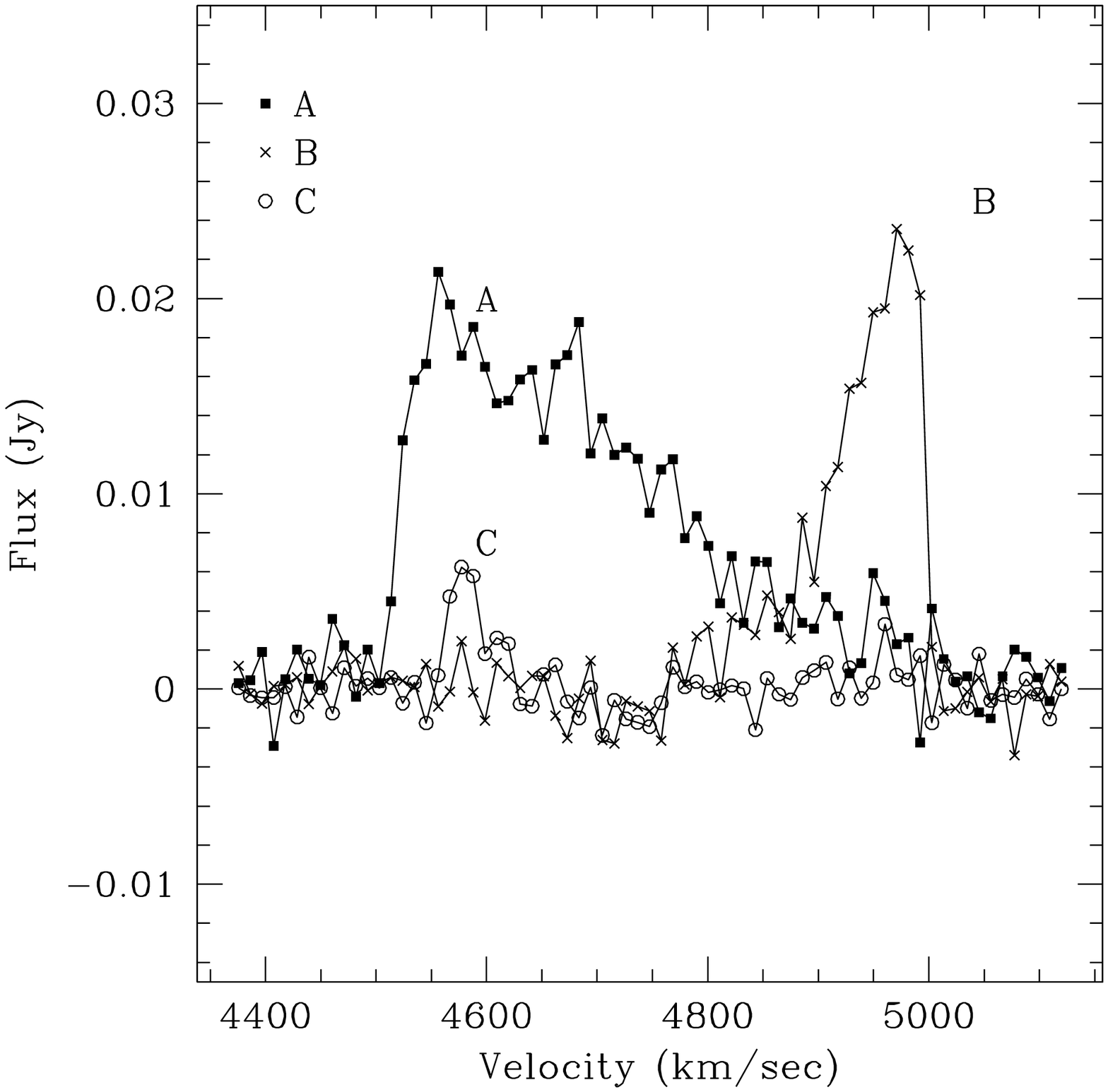}
\caption{Velocity profile of systems A, B, and C.\label{Fig8}}
\end{figure}


\begin{figure}
\figurenum{9a}
\plotone{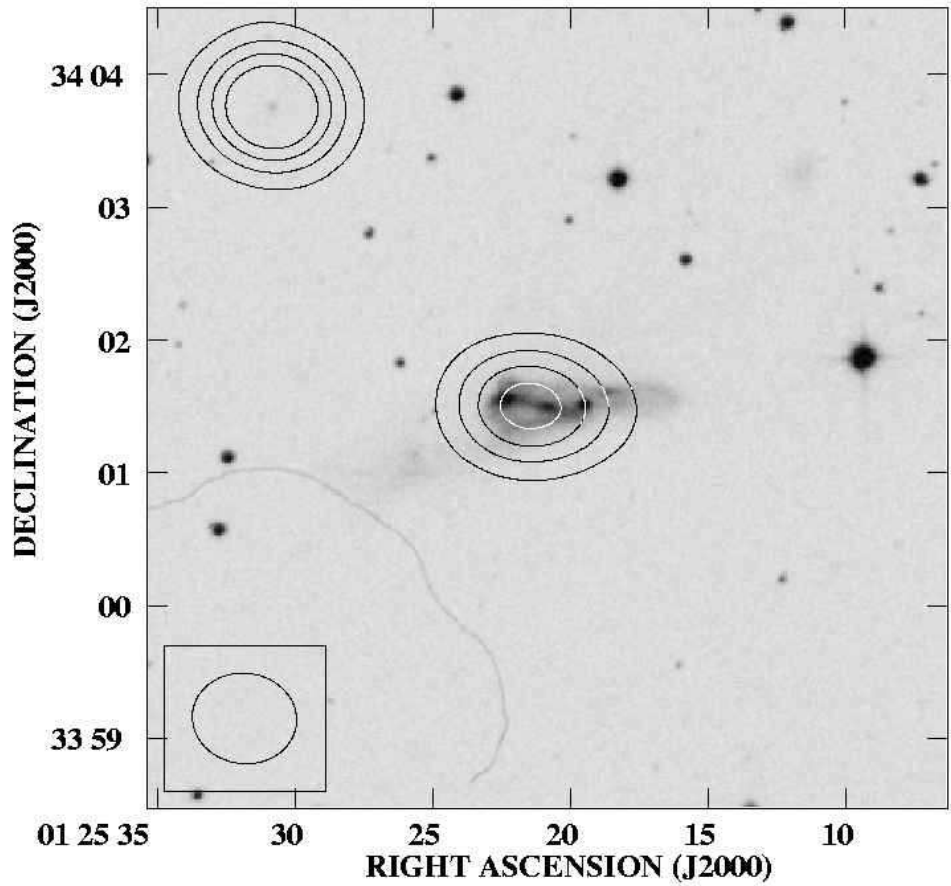}
\caption{The blue DSS2 optical image with integrated \HI\ continuum contours. The contours are drawn at (10, 20, 30, 40) $\times$ 0.26 mJy beam$^{-1}$ (0.26 mJy beam$^{-1}$ = 1 $\sigma$).\label{Fig9a}}
\end{figure}


\begin{figure}
\figurenum{9b}
\plotone{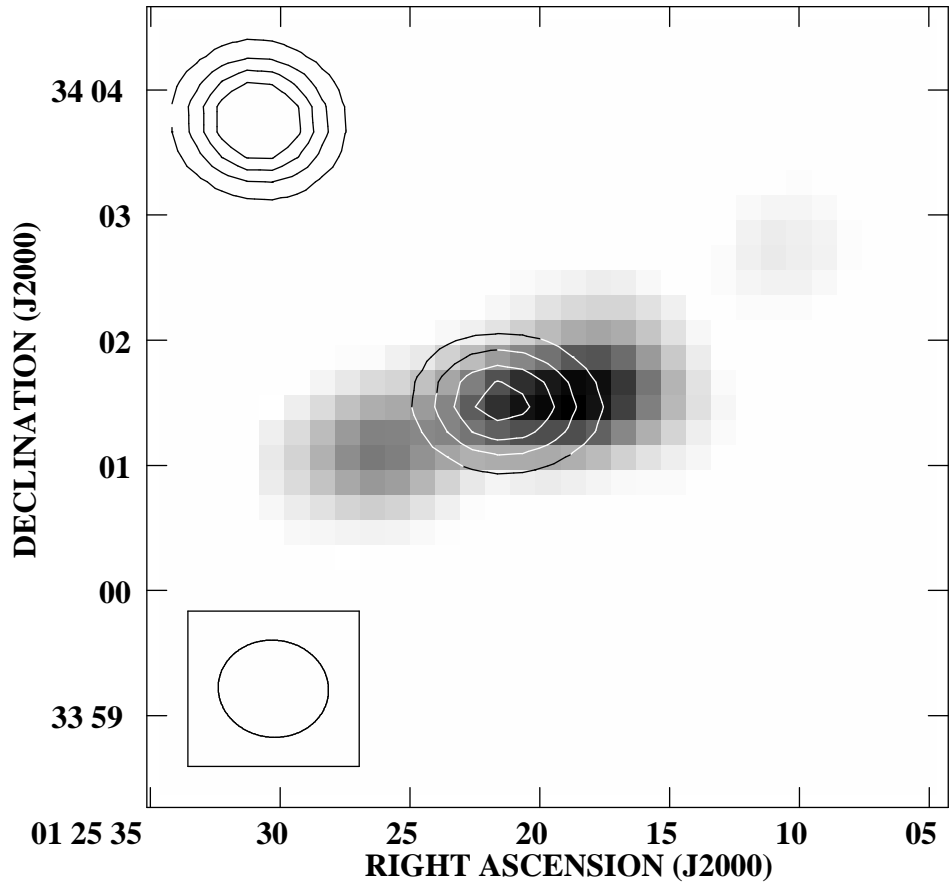}
\caption{Integrated \HI\ flux greyscale with \HI\ continuum contours. The contour levels are the same as in (9a).\label{fig9b}}
\end{figure}

\clearpage


\begin{deluxetable}{lc}

\tablewidth{0pt}
\tablenum{1}
\tablecaption{Properties of Arp 158\tablenotemark{a}\hspace{0.1in}\label{Table1}}
\tablehead{
\colhead{Parameter}&
\colhead{Value}
}

\startdata

NGC No. & 523\\
UGC No. & 00979 \\
VV No. & 783\\
Right Ascension (J2000) & 01$^{h}$ 22$^{m}$ 20.80$^{s}$ \\
Declination (J2000) &$ +34\arcdeg$ 01$\arcmin$ 29.0$\arcsec$ \\
Morphological Type & Pec\\
Distance\tablenotemark{b}\hspace{0.1in} (Mpc) & 63 \\
Photographic Magnitude  & 13.5 \\
Optical Dimensions (arcmin)  & 2.5 $\times$ $0.7$ \\

\enddata
\tablenotetext{a}{NASA IPAC Extragalactic Database}
\tablenotetext{b}{H$_0$ = 75 km s$^{-1}$ Mpc$^{-1}$}

\end{deluxetable}

\begin{deluxetable}{lc}
\tablewidth{0in}
\tablenum{2}
\tablecaption{VLA D-array Observations\label{Table2}}
\tablehead{
\colhead{Parameter}&
\colhead{Value}
}

\startdata

Date  & 14 and 17 May 1999 \\
No. of visibilities & 65344 \\
Bandwidth (MHz) & 6.25  \\
No. of channels & 127 \\
Channel Separation (kHz) & 48.8  \\
Heliocentric Velocity (km s$^{-1}$) & 4758  \\
Beam (FWHM) (arcsec) & 47.3 $\times$ 42.2 \\
rms noise in channel maps (mJy/Beam)  & 0.98  \\
Equivalent Brightness temperature (K) & 0.29 \\

\enddata
\end{deluxetable}

\begin{deluxetable}{lccccc}
\tablewidth{0in}
\tablenum{3}
\tablecaption{Derived Quantities\label{Table3}}
\tablehead{
\colhead{Object} &
\colhead {$\int$Sdv}&
\colhead{N$_{HI}$\tablenotemark{a}\hspace{0.1in}}&
\colhead{M$_{HI}$}&
\colhead{V$_{sys}$}&
\colhead{V$_{50}$}\\
\colhead{}&
\colhead {(Jy km s$^{-1}$)}&
\colhead{(atoms cm $^{-2}$)}&
\colhead {(M$_\odot$)}&
\colhead{(km s$^{-1}$)}&
\colhead{(km s$^{-1}$)}\\

}

\startdata

Arp 158 & 9.7 & \nodata  & 6.5 $\times$ 10$^{9}$& \nodata &\nodata\\
System A & 5.9 & 13.0 $\times$ 10$^{20}$ & 2.9$\times$ 10$^{9}$& 4614 & 256.4\\
System B & 2.4& 7.4 $\times$ 10$^{20}$ & 1.9 $\times$ 10$^{9}$ & 4957& 76.9\\
System C & 0.9 & 1.2 $\times$ 10$^{20}$ & 0.1 $\times$ 10$^{9}$& 4581 & 30.7\\
Tail & 2.6 & 8.1 $\times$ 10$^{20}$& 1.9 $\times$ 10$^{9}$& \nodata  &\nodata \\

\enddata
\tablenotetext{c}{The column density is the peak column density}
\end{deluxetable}

\end{document}